\documentclass[epsf]{aa}
\usepackage{graphicx}

\begin{document}
\def\fu{$f_1$}
\def\t{$\pm$}
\def\fd{$f_2$}
\def\ft{$f_3$}
\def\fq{$f_4$}
\def\fdu{$f_2 - 2f_1$}
\def\fp{$f_1 + f_2$}
\def\fm{$f_2 - f_1$}
\def\cd{cd$^{-1}$}
\def\cds{cd$^{-1}$\,}
\def\kms{km~s$^{-1}$}
\def\kmss{km~s$^{-1}$\,}
\def\I{\'\i}
\def\salp{\vskip 0.3truecm}
\title{HD~304373, the second case of 1O/2O double--mode Cepheid in
the Galaxy}
\author{M.~Beltrame\inst{1}, E.~Poretti\inst{2}}
\institute {Universit\`a degli Studi, Dipartimento di Fisica, Milano\\
\and
Osservatorio Astronomico di Brera, Via Bianchi 46,
I-23807 Merate, Italy\\
 \email{poretti@merate.mi.astro.it}
}
\offprints{E. Poretti}
\date{Received date; Accepted Date}
\abstract{We report on the discovery of the second case of a galactic Cepheid,
 ASAS~112843~--5925.7$\equiv$HD~304373, pulsating in the the first (1O) and second (2O) radial
overtones. The ratio between the periods (0.8058), the short value of the 1O period
(0.922405~d) and the shape of the 1O light curve makes HD~304373 very similar
to the 1O/2O Cepheids in the Magellanic Clouds. 
The implications of a so close similarity between a galactic 1O/2O pulsator and
LMC ones are discussed in terms of importance of the metallicity effects.
}
\authorrunning{Beltrame \& Poretti} 
\titlerunning{HD 304373}
\maketitle

\section{Introduction} 
The {\it All Sky Automated Survey} (ASAS) is a project started at the beginning of
1996; its ultimate goal is to detect any kind of photometric variability present
on a large area of the sky (Pojma\'nsky 1997). The aim has been pursued by
means of an automated mount equipped with a 716x512 pixel {\sc meade} CCD camera,
a 135-mm f/1.8 lens  and an $I$--band filter. The instrument is  located at Las 
Campanas Observatory in Chile and in the first three years it surveyed  
fifty 2x3~deg$^2$ fields. Detailed descriptions of the project,
the instrument, the data reduction and analysis can be found in three dedicated
papers (Pojma\'nsky 1997, 1998 and 2000).

In 2001 we started a careful re-examination of the pulsating stars included in the
ASAS catalogue, to find further confirmations of the characteristics of Cepheids
with $P<$8~d, RR Lyr stars  and
High Amplitude Delta Scuti stars.
Such a re-examination can be useful to detect new 
double--mode radial pulsators, which  are among the most elusive variable stars.
Their discovery constitutes a continuous observational challenge in the study
of stellar oscillations, especially in the Galaxy, where selection effects are 
very important. From a physical point of view, the period ratio directly allows us
to identify the pulsation modes. In turn, the matching of the observed period ratios
constitutes a powerful and practical tool to model the interior of stars. 
Pardo \& Poretti (1997) provide a systematic frequency analysis of the 
available photometry of the galactic double--mode Cepheids. Thirteen stars
show the fundamental ($F$) period ($P_0$) and the first overtone (1O, $P_1$) periods,
($P_1/P_0$ between 0.7127 and 0.6968), only one shows the 1O
and second (2O, $P_2$)  overtone periods (CO Aur, $P_2/P_1$=0.8008).  Therefore,
after the detection of a large number of double--mode Cepheids in the
Magellanic Clouds, it is quite evident that 
the discovery of new double--mode Cepheids,
and in particular new 1O/2O cases,  can help to compare
Cepheids of our Galaxy with those in other ones.

ASAS~112843~--5925.7 has been classified as a RR Lyr variable with
$P$=0.9220~d (Pojma\'nski 2000). The light curve has an amplitude of
0.22~mag in $I$--band.
It looks scattered, but no second periodicity has been suspected 
during the automatic period search, performed by analysis of
variance method (Pojma\'nski 1998). The {\sc simbad} database identifies ASAS 112843 --5925.7
as HD~304373$\equiv$TYC~8629-00990-1 (spectral type F8). 
{\sc tycho} photometry yields $B_T$=11.22$\pm$0.06 and $V_T$=10.57$\pm$0.05 
(H$\o$g et al. 2000).   
\section{Data analysis}
The original dataset consists of 544 measurements in the $I$--band. They span
about 1000 d, from 1997 April to 2000 January. The first 209 measurements were
collected in 40~days  in April and May 1997. Later, the survey started again in
October 1998 and lasted until January 2000, only stopped for the heliacal conjunction
(a gap of  70~day from August to October 1999). Thanks to this extensive 
coverage, the spectral window shows the
1~cy$^{-1}$ peak reduced at only 53\% of the power.  In many occasions HD~304373 was 
measured several times during every night and
therefore the alias at 1~\cds is also reduced (55\% of the power). Hence, the dataset
on HD~304373 appears very appropriate for an accurate frequency analysis.
In order to detect the frequency content of HD~304373 we applied the same technique
used by Pardo \& Poretti (1997). 
The least--squares power spectrum method (Vani\^cek 1971)
allowed us to detect one by one the constituent of the light curve. After each
detection, we refined the frequency values by applying the MTRAP code (Carpino et 
al. 1987), particularly suitable for the double--mode pulsation as it keeps
the relationships between
the detected terms (i.e., 2$f_1$, $f_1+f_2$,~...) locked. 

A quick glance to the original dataset was enough to discover the double--mode nature
of  HD~304373: two terms at $f_1$=1.084~\cds and $f_2$=1.345~\cds
clearly stand out, but a very high peak appeared at a very low frequency
in the subsequent power spectrum. Therefore, we repeated the analysis by subdividing 
the original datasets
into subsets (Tab.~\ref{lsq}) as the low--frequency peak suggested the presence of 
systematic errors.  In the first subset 
we detected not only the \fu~ and \fd~ term,
but also 2\fu~ and, more hidden in the noise, \fu+\fd. 
In the second dataset we detected again the 
\fu, 2\fu~ and \fd~ terms, but also
the \fu+\fd~ and 3\fu~ ones, which are clearly over the noise. In the third dataset 
 we only detected the \fu, 2\fu\, and \fd\, terms.
The second and third subsets had the same mean magnitude, while the first one showed a
remarkable systematic shift (Tab.~\ref{lsq}). Therefore, we built up
a homogenous dataset by shifting all the magnitudes of the first subset
by --0.029 mag.

\begin{figure}
\resizebox{\hsize}{!}{\includegraphics{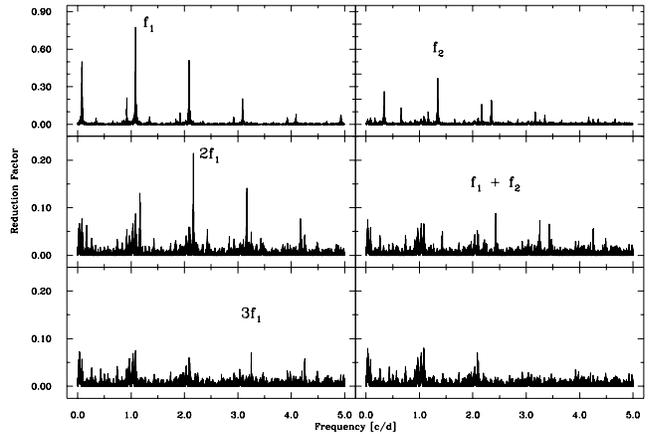}}
\caption[ ]{Power spectra of all the ASAS measurements of HD~304373. Each panel
shows the spectrum obtained by introducing all the previous identified frequencies as
known constituents; this means that the frequency values are considered
already determined, but their amplitude and phase values are worked out for each
new trial frequency. }
\label{power}
\end{figure}

The results of the frequency analysis of such a dataset is shown in 
Fig.~\ref{power}. The first power spectrum is dominated by the \fu=1.0842~\cds
term and its alias structure. This frequency is coincident with that
reported in the ASAS database. The low peak at \fd=1.3454~\cds 
becomes the highest in the second spectrum. The double--mode
nature of HD~304373 is clearly established looking at these two panels.
The subsequent analysis detected the 2\fu~ term (left panel in the middle row) and
the coupling term \fu+\fd~ (right panel in the middle row). The presence
of the coupling term ruled out the possibility of a binary system composed
of two pulsating stars. In the bottom row the signal detection is more complicated.
If the 3\fu~ term can be recognized in the fifth panel, the sixth one
looks characterized by a bunch of frequencies around integer values of ~\cd.
One peak structure is at the sinodical month ($f$=0.034~\cd) and its aliases, 
reflecting the Full--Moon interference: this spurious peak is quoted
by Pojma\'nski (1998) as a common result in his period search in the
ASAS database. Another one is close to the \fu~ value, just at the
limit of the frequency resolution. Its presence can be explained pointing out 
that the amplitude of the $f_1$ term results smaller in the third dataset
(0.067$\pm$0.003 mag) than in the other two (0.089$\pm$0.002, 0.090$\pm$0.002 mag).
This term disappears analyzing only the first two datasets combined together. 
We also modified the last 90 measurements by amplifying the contribution of
the first frequency, increasing its amplitude by a factor 1.33. In such a way,
we obtained a dataset in which the $A_1$ term is constant. The frequency
analysis detected the $f_1, f_2, 2f_1, f_1+f_2, 3f_1$ and $f$=0.034~\cd terms,
 but no peak close to $f_1$. Therefore, it is definitely established
that the smaller $A_1$ amplitude in the third dataset is responsible for the
observed doublet.

\section{Least--squares fits and Fourier decomposition}

Table~\ref{lsq} reports the parameters of the least--squares fits obtained
by means of the formula
\begin{equation}
I(t)= I_o + \sum_z {A_z \cos [2\pi f_z  (t-T_o) +\phi_z ]}
\end{equation}
where $f_z$ is the generic frequency, which can be an independent frequency
(\fu and \fd), a harmonic (2\fu and 3\fu) or the cross coupling term \fu+\fd.
The fits of the three subsets are reported for comparison purposes; the procedure
to identify the terms present in the light curve of HD~304373 is the same used by
Pardo \& Poretti (1997, see their Tab.~3), but here we failed to build a homogenous
dataset owing to the differences in amplitudes.
Indeed, the amplitude variation of $f_1$ looks big; its instrumental origin is not confirmed
by the values of the \fd~ term, which remains constant
among the three
subsets. Moreover, the amplitude ratio $R_{21}$ (0.18$\pm$0.02, 0.22$\pm$0.02 and 0.18$\pm$0.04)
and the phase difference $\phi_{21}$ 
(3.89$\pm$0.20, 3.78$\pm$0.18, 3.62$\pm$0.33 rad)
of the \fu~term  remain constant. That does 
not support a beating modulation between two close frequencies 
as in this case the shape of the \fu~ light curve should vary
following the beat phase. 
Our analysis clearly indicates a sudden decrease of the $f_1$ amplitude, but it
is not able to point out the reason precisely. Therefore, we considered the 454 measurements
between JD 2450546 and 2451402 as the final dataset, adjusted for the systematic differences.

\begin{figure}
\resizebox{\hsize}{!}{\includegraphics{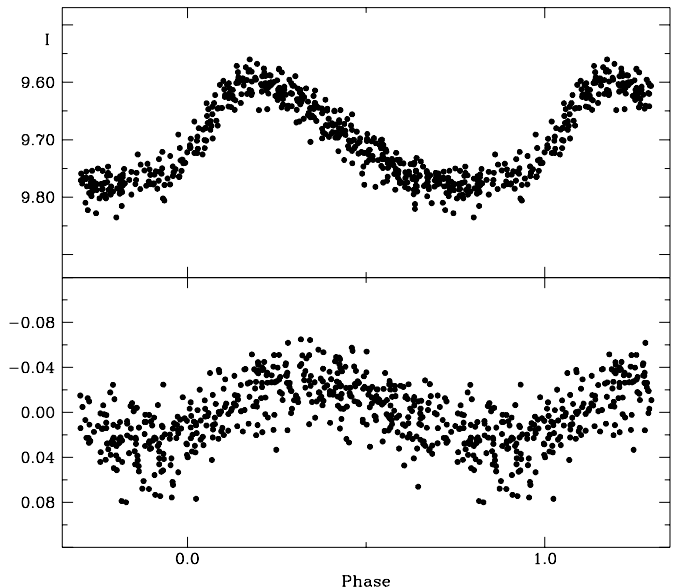}}
\caption[ ]{Light curves of the two independent frequencies \fu=1.084123~\cds
(upper panel) and \fd=1.345458~\cds (lower panel) as obtained from the measurements
of HD~304373 between JD 2450546 and 2451402.}
\label{curves}
\end{figure}

In order to fit the \fu~ 
curve three harmonics are necessary, while the \fd~ curve does not require
any harmonics. Therefore it is possible to calculate the Fourier parameters
only in the former case. The values obtained from the final 
dataset are  $\phi_{21}=3.80\pm$0.11 rad, $\phi_{31}=1.75\pm$0.23 rad and
$R_{21}$=0.21$\pm$0.02. 
Figure~\ref{curves} shows the light curves of the two periods of HD~304373: 
the upper curve was obtained by subtracting the \fd~ and \fu+\fd~
terms from the measurements, the lower one by
subtracting \fu, 2\fu, 3\fu, \fu+\fd.
Note also that the spurious term at $f$=0.034~\cds 
was not removed from the data. When considering it, the least--squares
parameters don't change in a significant way and the residual rms 
decreases just a little, to 0.019 mag.

\begin{table*}
\caption{Parameters of the least--squares fits of the ASAS measurements on
HD~304373. They are divided into two subsets owing to the different mean
magnitudes and to the change in the amplitude of the \fu term. $T_0$=HJD~2451026.2671}
\begin{tabular} {l c r  c  rr c rr c rr  c rr }
\hline
& & & & \multicolumn{2}{c}{JD 2450546-2450584}& & \multicolumn{2}{c}{JD 2451109-2451402} & &
\multicolumn{2}{c}{JD 2451469-2451583}& & \multicolumn{2}{c}{JD 2450546-2451402}\\
\cline{5-6}\cline{8-9}\cline{11-12}\cline{14-15}
\multicolumn{1}{c}{Term} & & \multicolumn{1}{c}{Freq.} & &\multicolumn{1}{c}{Ampl.} &
\multicolumn{1}{c}{Phase} & &\multicolumn{1}{c}{Ampl.} & \multicolumn{1}{c}{Phase} & &\multicolumn{1}{c}{Ampl.} &
\multicolumn{1}{c}{Phase}& & \multicolumn{1}{c}{Ampl.} & \multicolumn{1}{c}{Phase} \\
\multicolumn{1}{c}{} & & \multicolumn{1}{c}{[\cd]} & & \multicolumn{1}{c}{[mag]} & \multicolumn{1}{c}{[rad]} & &
\multicolumn{1}{c}{[mag]} & \multicolumn{1}{c}{[rad]}& &
\multicolumn{1}{c}{[mag]} & \multicolumn{1}{c}{[rad]}& & \multicolumn{1}{c}{[mag]} & \multicolumn{1}{c}{[rad]} \\
\hline
$f_1$ & & 1.084123 & & 0.089 & 5.82 & & 0.090 &  5.84& &  0.067 &  5.73 & &  0.089 & 5.83\\
 & & $\pm$.000007 & & $\pm$.002 & $\pm$.03 & & $\pm$.002 & $\pm$.02 & &$\pm$.003 & $\pm$.05 & & $\pm$.002 & $\pm$.02\\ 
$f_2$ & & 1.345458 & &    0.028 &     2.11 & &     0.027 &     2.06 & &    0.029 &     2.11 & &     0.028 &     2.08\\
& & $\pm$.000021 & & $ \pm$.002 & $\pm$.08 & & $\pm$.002 & $\pm$.06 & &$\pm$.003 & $\pm$.10 & & $\pm$.002 & $\pm$.05\\
 $2f_1$& &         & &    0.016 &     2.97 & &     0.020 &     2.90 & &    0.012 &     2.52 & &     0.019 &     2.90\\
      & &          & &$\pm$.002 &$ \pm$.14 & & $\pm$.002&  $\pm$.08 & &$\pm$.003 &$ \pm$.23 & & $\pm$.002&  $\pm$.07\\ 
$f_1+f_2$& &       & &    0.010 &     6.12 & &     0.011  &    6.10 & &     --   &  --      & &     0.010 &     6.09\\
      & &          & &$\pm$.002 &$ \pm$.23 & & $\pm$0.002& $\pm$.16 & &          &          & & $\pm$.002 &$ \pm$.13\\ 
$3f_1$ & &         & &     --   &      --  & &     0.019 &     0.75 & &     --   &  --      & &     0.008 &     0.40\\
      & &          & &          &          & & $\pm$0.002& $\pm$.18 & &          &          & & $\pm$.002&  $\pm$.17\\
\noalign{\smallskip}

\multicolumn{3}{c}{Mean $I$ magnitude} &&\multicolumn{2}{c}{9.734$\pm$.002} &
& \multicolumn{2}{c}{9.705$\pm$.001}&& \multicolumn{2}{c}{9.704$\pm$.002} & 
& \multicolumn{2}{c}{9.705$\pm$.002}\\
\multicolumn{3}{c}{Residual r.m.s. [mag]} &&\multicolumn{2}{c}{0.022} & & \multicolumn{2}{c}{0.019}&& \multicolumn{2}{c}{0.019} & & \multicolumn{2}{c}{0.020}\\
\multicolumn{3}{c}{N}&&\multicolumn{2}{c}{209} && \multicolumn{2}{c}{245}& &\multicolumn{2}{c}{90} && \multicolumn{2}{c}{454}\\
\hline
\end{tabular}
\label{lsq}
\end{table*}

\section{The double--mode pulsation}
Our analysis detected the simultaneous excitation of two modes in the light
curve of HD~304373: $P_1$=0.922405~d and $P_2$=0.743241~d. The resulting
ratio, $P_2/P_1$=0.8058, is that expected for Cepheids pulsating in the 1O
and 2O radial modes.  In the Galaxy there was only one similar star so far,
i.e., CO Aur ($P_1$=1.78303~d, $P_2$=1.42778~d, $P_2/P_1$=0.8008; Mantegazza 1983).

\begin{figure}
\resizebox{\hsize}{!}{\includegraphics{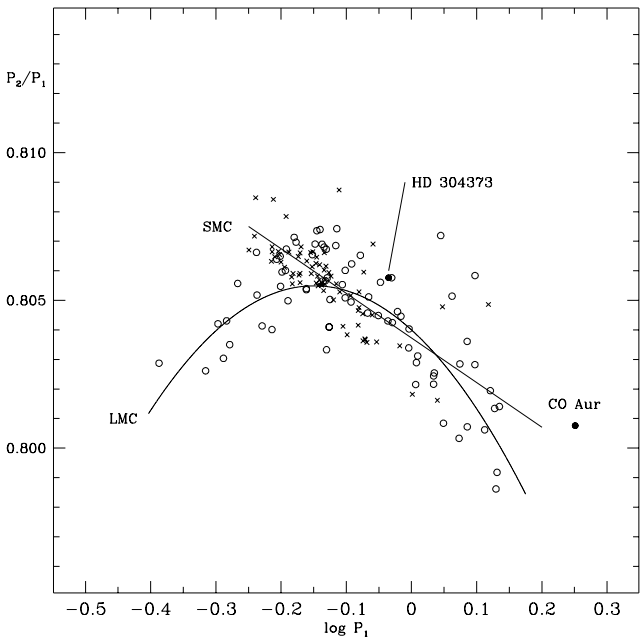}}
\caption[ ]{Period ratios among the 1O/2O pulsators in the Small (crosses) and
Large (circles) Magellanic Clouds. Fits are also shown. The filled circles
indicate the value observed for the galactic 1O/2O pulsators, HD~304373 and
CO Aur.}
\label{fs}
\end{figure}

Figure~\ref{fs}
shows the $P_2/P_1$ ratio as a function of $\log~P_1$ period (Petersen diagram)
for 1O/2O pulsators
in the Large (Alcock et al. 1999; Soszy\'nski et al. 2000) and in the Small
(Udalski et al. 1999)
Magellanic Clouds. The parabolic fit of the LMC stars shown in Fig.~\ref{fs} 
is after Soszy\'nski et al. (2000), while the linear fit of the SMC 
stars has been worked out again. 
Figure~\ref{fs} gives evidence that the $P_2/P_1$ values are not well
separated in the Magellanic Clouds. It implies that the $P_2/P_1$ ratio is
less sensitive than the $P_1/P_0$ ratio to the difference in metallicity
between the Magellanic Clouds.   
Indeed, the $P_1/P_0$  ratio for each galaxy defines three well separated
lines in the Petersen diagram (see Fig.~1 in Soszy\'nski et al. 2000). 

The $P_2/P_1$ value related to HD~304373
is quite normal for both samples and  
very similar to that of {\sc macho}*05:30:11.7 -69:52:02, a star 
belonging to the LMC (Fig.~\ref{fs}). Moreover, this ratio cannot be
matched by the galactic composition models proposed by Morgan \& Welch
(1997), while it can be by the SMC and LMC ones.
If HD~304373 is a normal Pop.~I star, its $P_2/P_1$ value 
suggests a small dependence on metallicity, as it does not change in
a environment as the Milky Way, which is more metallic than the Clouds.

We obtain $<I>$=16.51 (standard deviation $\pm$0.12 mag) considering the eleven
1O/2O Cepheids with 0.80$<P_1<$1.00~d in the LMC (see Tab.~2 in 
Soszy\'nski et al. 2000). Assuming $m-M$=18.5 for
LMC stars, we get  $<M_I>$=--2.0 for 1O/2O Cepheids. Therefore,
$I$=9.70 locates  HD~304373  at 2.2~Kpc from the Sun. 
Taking into account the galactic coordinates ($l=293^o$,
$b=2^o$), HD~304373 is 77~pc away from the galactic plane. These
considerations support the hypothesis that HD~304373 is a disk
Pop.~I star. Christensen-Dalsgaard \& Petersen (1995) pointed out
that the matching between the F/1O ratios for galactic pulsators
and the theoretical models occurs for metallicities smaller than
the solar value of 0.017--0.020; if that applies for the 1O/2O
pulsators too, a metallicity close to 0.010 allows the $P_2/P_1=0.8058$
and $P_1$=0.922405~d values to reasonably fit the theoretical
models (Christensen-Dalsgaard \& Petersen 1995; Morgan \& Welch 1997).

Moreover, we notice that the
scatter of the $P_2/P_1$ values observed in both 
Clouds is intrinsic, i.e., originated from slightly different physical conditions
inside the stars (it is much larger than the error bars on the period ratios,
typically a few units of 10$^{-5}$, since the periods are known with high accuracy). 
Therefore, the weak effect of the metallicity can be masked by other reasons
(see also Fig.~4 in Christensen-Dalsgaard \& Petersen 1995).

As in the case of CO Aur, there is no significant contribution of the 2\fd\,
harmonic in the light curve of HD~304373, i.e., it is perfectly sine--shaped
within error bars. This is quite common among the 1O/2O pulsators.
On the other hand, the $\phi_{21}$, $\phi_{31}$ and $R_{21}$ values found
for the 1O light curve of HD~304373 are in excellent agreement with the values
observed in the 1O light curves of 1O/2O pulsators in the Magellanic Clouds.

\section{Conclusions}
In each class (HADS, RR Lyr and Cepheids) the shortest period stars  are typical
of metal--poor environments. The more evident cases are the very--short period
SX Phe stars observed in globular clusters. 
HD~304373 has a very short fundamental period for
a classical Cepheid: assuming $P_1/P_0$=0.70, we obtain $P_0$=1.32~d. This
result supports  the previous discussion, suggesting  a low metallicity for
HD~304373 and  explaining why 
its $P_2/P_1$ value is so  similar to that of  1O/2O pulsators in the Magellanic Clouds.

From a methodological point of view, 
we identified the presence of a peak close to $f_1$ as the result of a smaller
amplitude in the last observing season. No reasonable physical explanation has been
found and its instrumental origin is likely.
We also detected the influence of  a non perfect
sky subtraction or flat--fielding in presence of the Full Moon.
It is interesting to note that this instrumental
effect makes HD~304373 brighter at the New Moon and fainter at the Full Moon.  
We stress the
importance of checking the homogeinity of the time--series on pulsating stars
before processing them.

\begin{acknowledgements}
The research has made use of the {\sc simbad} database, operated at CDS, Strasbourg,
France.
The authors wish to thank L.~E.~Pasinetti for help and encouragement in
this work, E.~Antonello, L.~Mantegazza, S.M.~Morgan, J.O.~Petersen, L.~Szabados and
the referee, P.~Moskalik, 
for useful comments on a first draft of the manuscript. 
\end{acknowledgements}

\end{document}